\documentclass[prl,aps,amssymb,twocolumn,showpacs,floatfix]{revtex4}

\usepackage{graphics}

\usepackage{epsfig} 

\usepackage{dcolumn}

\usepackage{amsmath}

\begin{document}

\newcommand\dd{{\operatorname{d}}}
\newcommand\sgn{{\operatorname{sgn}}}
\def\Eq#1{{Eq.~(\ref{#1})}}
\def\Ref#1{(\ref{#1})}
\newcommand\e{{\mathrm e}}
\newcommand\cum[1]{  {\Bigl< \!\! \Bigl< {#1} \Bigr>\!\!\Bigr>}}
\newcommand\vf{v_{_\text{F}}}
\newcommand\pf{p_{_\text{F}}}
\newcommand\ef{{\varepsilon} _{\text{\sc f}}}
\newcommand\zf{z_{_\text{F}}}
\newcommand\zfi[1]{{z_{_\text{F}}}_{#1}}
\newcommand\av[1]{\left<{#1}\right>}

\title{Non-equilibrium Luttinger liquid: Zero-bias anomaly and dephasing}

\author{D. B. Gutman$^{1,2}$, Yuval Gefen$^3$, and A. D. Mirlin$^{4,1,2,5}$}
\affiliation{\mbox{$^1$Institut f\"ur Theorie der kondensierten Materie,
 Universit\"at Karlsruhe, 76128 Karlsruhe, Germany}\\
\mbox{$^2$DFG--Center for Functional Nanostructures,
 Universit\"at Karlsruhe, 76128 Karlsruhe, Germany}\\
\mbox{$^3$Department of Condensed Matter Physics, Weizmann Institute of
  Science, Rehovot 76100, Israel}\\
\mbox{$^4$Institut f\"ur Nanotechnologie, Forschungszentrum Karlsruhe,
 76021 Karlsruhe, Germany}\\
\mbox{$^5$ Petersburg Nuclear Physics Institute, 188300
  St. Petersburg, Russia}
}

\date{\today}

\begin{abstract}
A one-dimensional system of interacting electrons out of equilibrium is
studied in the framework of the Luttinger liquid model.
We analyze several setups and develop a theory of
tunneling into such systems. A remarkable property of the problem is
the absence of relaxation in energy distribution functions of left-
and right-movers, yet the presence of the finite dephasing rate
due to electron-electron scattering, which
smears zero-bias-anomaly singularities in the tunneling density
of states. 
\end{abstract}

\pacs{73.23.-b, 73.40.Gk, 73.50.Td
\\[-0.5cm] 
}


\maketitle 

The interest in one-dimensional (1D) interacting electron systems is
due to their fascinating physical
properties and
potential applications in \mbox{nanoelectronics}. A variety of
experimental realizations of quantum wires includes carbon
nanotubes, quantum Hall edges, semiconductor structures, 
polymer fibers, and metallic nanowires. 
At equilibrium,  the physics of 1D electrons  
has been thoroughly explored.
The main peculiarity of the system is the formation of a strongly
correlated state, Lut\-tin\-ger liquid (LL), commonly described in terms
of collective bosonic excitations \cite{gnt}. A hallmark of
LL correlations is a strong, power-law suppression of tunneling
current at low bias --- zero-bias anomaly (ZBA) \cite{zba-carbon-nanotubes}.

On the other hand, little is known about LL away from equilibrium.
Theoretical efforts so far focused on non-linear
transport through a single impurity \cite{Saleur,Feldman}.
In this work we consider the problem of tunneling into a LL out of
equilibrium. It is important to emphasize a remarkable property of the
LL: the absence of relaxation towards equilibrium. Once non-equilibrium
energy distributions of left- and right-movers are created, they 
propagate along the wire arbitrarily long without thermalization (within
the clean LL model, i.e. when back-scattering by interaction or
impurities, spectral nonlinearity, and momentum dependence of
interaction are neglected) \cite{energy_relaxation}. 
Below we address such a situation and analyze the tunneling density of
states (TDOS) of a non-equilibrium LL wire. We show that while the
energy relaxation rate is zero, the ZBA in TDOS is smeared out by
dephasing processes yielding a finite quasiparticle life time. 

We begin by discussing possible experimental realizations of a 
non-equilibrium LL. Throughout most of the paper we assume 
that the interaction strength interpolates
adiabatically between its value in the LL (central part of the wire
where the measurements are performed) and zero near the electrodes. 
The assumption of adiabaticity is not
entirely innocent; at the end, we briefly discuss the case of  
sharp switching of interaction and ensuing modifications. 

\begin{figure}[htbp]
\includegraphics[width=0.8\columnwidth,angle=0]{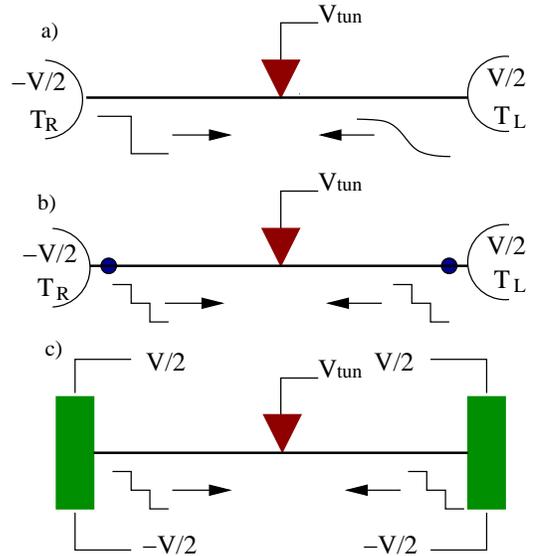}
\caption{Schematic view of setups for measurement of the
  TDOS of a LL out of
  equilibrium (see text for details).}
\label{fig1}
\end{figure}

The simplest setup is shown in Fig.~\ref{fig1}a.
A long clean LL is adiabatically coupled to two electrodes 
with different potentials, $\mu_L - \mu_R = eV$ 
and different
temperatures $T_\eta$ (where $\eta=L,R$ stands for left-
and right-movers) \cite{Chudnovsky}.  A particularly interesting
situation arises when one of temperatures is much 
larger than the other, e.g., $T_L=0$
and $T_R$ finite. Then the ZBA at $\mu_L$, where the distribution
function has a sharp step, is broadened solely by dephasing
originating from electron-electron scattering. 

A more complex situation arises if electrons coming from reservoirs
with different potentials mix by impurity scattering.
Two different realizations of such devices are shown in
Figs.~\ref{fig1}b,c.  
In the first case, Fig.~\ref{fig1}b, the mixture of left and right
movers coming from reservoirs with $\mu_L \ne \mu_R$ 
is caused by 
impurities which are located in the  non-interacting part of the wires. 
In the second setup, Fig.~\ref{fig1}c, the LL wire is
attached to two metallic wires which are themselves
biased. We assume that these electrodes are diffusive but 
sufficiently short, so that inter-electrode energy equilibration can be
neglected. As a result, a double-step energy
distribution is formed in each electrode \cite{Pothier}. 
The left- and right- movers in the LL
wire ``inherit'' these non-equilibrium distributions emanating from
the respective electrodes. 
The existence of multiple Fermi edges in the  distribution
functions ``injected'' from the electrodes renders the behavior 
of the TDOS highly non-trivial, since the ZBA is expected to be broadened by 
electron-electron scattering processes \cite{GGM} governing
the dephasing rate $\tau_\phi$. 

The question of non-equilibrium ZBA broadening induced by
electron-electron scattering is
particularly intriguing in the case of a 1D system. 
First, energy relaxation is absent in that case. Second, 
there are two qualitatively different predictions concerning dephasing
in the context of weak localization and
Aharonov-Bohm oscillations: while the weak-localization dephasing rate
vanishes in the limit of vanishing disorder \cite{GMP}, 
the Aharonov-Bohm dephasing rate is finite in a clean LL
\cite{GMP,lehur}. It is thus very interesting to see how
dephasing processes manifest themselves in the broadening of ZBA.
From the technical point of view, the challenge is to develop 
methods to treat LL away from equilibrium.

Within the LL model, the electron field is decoupled into the sum of left-
and right-moving terms,  
$\psi(x,t)=\psi_R(x,t)e^{ip_Fx}+\psi_L(x,t)e^{-ip_Fx}$. The
Hamiltonian reads
$$
H=iv \left(\psi_R^\dagger\partial_x\psi_R-
\psi^\dagger_L\partial_x\psi_L\right)+\frac{V_0}{2}(\psi_R^\dagger\psi_R
+ \psi_L^\dagger\psi_L)^2,
$$
where $V_0$ is the bare electron-electron interaction.
We will neglect the interaction between the tip and the wire,
assume that the tunneling current is weak and that electrons in the
tip are at equilibrium at a negligibly low temperature. We further
assume that the TDOS of the tip, $\nu_t$, can be considered as constant. 
Then, dependence of the differential tunneling conductance on the
voltage $V_t$ at the tip
is controlled by the TDOS of the wire 
\begin{equation}
\label{diff-cond}
\partial I / \partial V_t \propto |T|^2\nu_t\nu(eV_t)\ ,
\end{equation}
where $T$ is a tunneling matrix element. We thus focus on 
the TDOS $\nu(\epsilon)$ of a non-equilibrium LL.

In order to find the TDOS $\nu(\epsilon) = \nu_R(\epsilon) +
\nu_L(\epsilon)$ at a point $x$
one  needs to calculate the retarded (``r'')
and advanced  (``a'') single particle Green function,
$\nu_\eta(\epsilon)={i\over
  2\pi}[G^r_\eta(\epsilon,x,x)-G^a_\eta(\epsilon,x,x)]$.
To do it away from equilibrium  
we employ the Keldysh formalism. The Keldysh Green function reads 
$G_{\eta}(x,t;x',t')=-i\left\langle T_K \psi_{\eta}(x,t)\psi_{\eta}^\dagger 
(x',t')\right\rangle$,
where the time ordering is along the Keldysh contour.  
We proceed by following the lines of functional bosonization approach
\cite{Fogedby:76,LeeChen:88,Yur}. While fully equivalent to the
conventional bosonization 
technique for the case of clean equilibrium LL, this method is
advantageous in more complicated situations. In particular, the
efficiency of the functional bosonization for the analysis of
transport and interference phenomena in a disordered LL was recently
demonstrated \cite{GMP}. The key features of the functional
bosonization 
are: 
(i) it retains explicitly both fermionic and bosonic
degrees of freedom; (ii) for $V_0=0$ it
straightforwardly reduces to the model of non-interacting fermions.  
This makes the functional bosonization an appropriate tool for the development
of the theory of a non-equilibrium LL.

Decoupling the
interaction by Hubbard-Stra\-to\-no\-vich transformation 
via a bosonic field $\phi$, we obtain the action 
\begin{align}\label{TL}
S[\psi,\phi]=i\sum_{\eta=R,L}\psi^*_\eta
(\partial_\eta-\phi)\psi_\eta-\frac{1}{2}\phi 
V_0^{-1}\phi\, ,
\end{align}
where $\partial_{R,L} = \partial_t\pm v\partial_x$.
 It is convenient to  
perform a rotation in Keldysh space, thus
decomposing fields into classical and quantum components,
$\psi,\bar{\psi} = (\psi_+ \pm \psi_-)/\sqrt{2}$ 
(where $+$ and $-$ label the fields  on two
branches of the contour)
and analogously for
$\phi$. We further introduce vector notations by combining $\phi$
and $\bar{\phi}$ in a 2-vector $\vec{\phi}$. 

The Green function of interacting electrons can be presented in the form
$G_\eta(t-t',x-x')= \langle G_{\eta\phi}(x,t;x',t')\rangle$,
where $G_{\eta\phi}$ is the Green function of non-interacting fermions in an
external field $\phi$, the averaging goes with the weight 
$Z_\phi e^{-\frac{i}{2}\vec{\phi}^T V_0^{-1}\sigma_1\vec{\phi}}$, and
$Z_\phi$ is a sum of vacuum diagrams (fermionic loops) in the field $\phi$.
The special feature of 1D geometry is that the coupling between the
fermionic and bosonic fields can be eliminated by a gauge
transformation, 
$\nolinebreak{\psi_{\eta}(x,t)\to\psi_{\eta}(x,t)e^{i\Theta_{\eta}(x,t)}}$,
with $\Theta_\eta = \sigma_0\theta + \sigma_1\bar{\theta}$, if we require
\begin{eqnarray}&&
\label{diff_eq}
i\partial_{\eta}\vec{\theta}_{\eta}=\vec{\phi}\,.
\end{eqnarray}
As a result, $G_{\eta\phi}$ can be cast in the form
\begin{eqnarray}&& 
\hspace{-0.5cm}G_{\eta\phi}(x,t;x',t')=e^{i\Theta_\eta(t)}G_{\eta 0}(x,x';t-t')
e^{-i\Theta_\eta(t')}\,, \nonumber \\ 
&&G_{\eta 0}=
\begin{pmatrix}
     G^r_{\eta 0} & G^K_{\eta 0}\\
      0 & G^a_{\eta 0}
    \end{pmatrix}\:\!.
\label{green-function-free} 
\end{eqnarray}
Here $G_{\eta 0}$ is a Green function of free fermions, with the
Keldysh component $G_{\eta 0}^K(\epsilon) = 
[1-2n_\eta(\epsilon)][G_{\eta 0}^r(\epsilon)-G_{\eta 0}^a(\epsilon)]$, and
$n_\eta(\epsilon)$ is fermionic distribution function.
  
To proceed further, we use the random-phase approximation (RPA),
within which $Z_\phi$ is Gaussian, 
$\nolinebreak{\log Z_\phi=-\frac{i}{2}\vec{\phi}^T\Pi\vec{\phi}}$. 
This is an exact relation at equilibrium \cite{DzLar:73},
which is crucial for the LL being an exactly solvable problem. 
It remains exact for the non-equilibrium setup of Fig.~\ref{fig1}a,
where both distributions $n_\eta$ are of Fermi-Dirac form. On the
other hand, RPA becomes an approximation for more general
non-equi\-lib\-ri\-um situations (cf. Figs.~\ref{fig1}b,c); we 
discuss its status and possibility of an exact solution at the end. 

The polarization operator of free fermions is given by 
$\Pi=\Pi_R+\Pi_L$, with
\begin{eqnarray}&&
\label{J5}
\Pi^r_{R,L}=-\frac{1}{2\pi}\frac{q}{\omega_+\mp v_Fq}\,\, , \,\,
\Pi^a_{R,L}=-\frac{1}{2\pi}\frac{q}{\omega_-\mp v_Fq}\, ,  \nonumber \\&&
\Pi^K_\eta =(\Pi^r_\eta-\Pi^a_\eta)B^v_\eta(\omega)\,,
\end{eqnarray}
where $\omega_\pm =\omega\pm i\delta$. 
The function
\begin{equation}
\label{e7}
B^v_\eta(\omega)=\frac{2}{\omega}\int_{-\infty}^{\infty} d\epsilon\:
n_\eta(\epsilon)\:[2-n_\eta(\epsilon-\omega)-n_\eta(\epsilon+\omega)],
\end{equation}
is related to the distribution function $N_\eta^v(\omega)$ of electron-hole 
excitations  moving with velocity $v$ in direction $\eta$, 
$B^v_\eta(\omega) = 1 + 2 N^v_\eta(\omega)$.  At equilibrium,
$B^v_\eta(\omega) = B_{\rm eq}(\omega) = 1+ 2N_{\rm eq}(\omega)$, where
$N_{\rm eq}(\omega)$ is the Bose distribution.
Since fermions are free after the gauge transformation, no relaxation 
of the distribution functions $n_\eta(\epsilon)$ takes place.

Performing the averaging over $\phi$, we express 
the TDOS in terms of the correlation function of the gauge fields   
\begin{eqnarray}
\label{J6}
\frac{2\nu_\eta(\epsilon)}{\nu_0}=1+2i\int_{-\infty}^{\infty}dt n_\eta(t)\exp\left(-I_{\theta\theta}^{(\eta)}\right)\sin(I_{\theta\bar{\theta}}^{(\eta)})\,.
\end{eqnarray}
Here  $\nu_0$ is the bare (non-interacting) density of states,  $n_\eta(t)$ is
the Fourier transform of $n_\eta(\epsilon)$,  
\begin{eqnarray}&&
I_{\theta\theta}^{(\eta)}(t)=\int (d\omega)(dq)(1-\cos(\omega t))\langle \theta \theta\rangle_{\omega,q}^{(\eta)}e^{-|\omega|/\Lambda}\,, \nonumber \\&&
I_{\theta\bar{\theta}}^{(\eta)}=2\int (d\omega)(dq)\sin(\omega
t)\langle\theta\bar{\theta}\rangle_{\omega,q}^{(\eta)}e^{-|\omega|/\Lambda}\, ,
\label{J6a}
\end{eqnarray}
and $\Lambda$ is an ultraviolet cutoff.
To calculate $I_{\theta\theta}$ and $I_{\theta\bar{\theta}}$ one needs
to resolve Eq.~(\ref{diff_eq}) and to
express the gauge field $\vec{\theta}$ in terms of the
Hubbard-Stratonovich filed $\vec{\phi}$,
\begin{align}
\label{aa1}
\vec{\theta}_\eta={\cal G}_{\eta 0}\vec{\phi} \,,
\end{align}
where ${\cal G}_{\eta 0}$ is the Green function of free bosons with the
Keldysh component
${\cal G}_{\eta 0}^K = ({\cal G}_{\eta 0}^r-{\cal G}_{\eta 0}^a)B^v_\eta$. 
The latter serves to reproduce correctly the distribution
function $N_\eta^v$ in the $\langle\theta\theta\rangle$ correlation
function, ensuring, in particular, the fluctuation-dissipation
theorem at equilibrium.

The correlation functions of $\vec{\theta}$ fields can be now readily found. 
For the $\langle \theta \bar{\theta}\rangle$
component (which is independent of the distribution functions) we get 
\begin{equation}
\label{J8}
\langle \theta \bar{\theta}\rangle_{\omega,q}^{(R,L)}
=\frac{1}{2}\frac{\langle \phi
  \bar{\phi}\rangle}{(\omega_+\mp v q)^2}=
  \frac{iV_0}{2}\frac{\omega\pm v q}{(\omega_+\mp vq)(\omega_+^2-u^2q^2)},\nonumber
\end{equation}
where $u = v(1+V_0/\pi v)^{1/2}$ is the sound velocity.
This yields for the $q$-integrated propagator [which enters
Eq.~(\ref{J6a})]
$\int (dq) \langle \theta \bar{\theta}\rangle_{\omega q}^{(\eta)}
= \pi\gamma/2\omega $,
where   $\nolinebreak{\gamma=(K-1)^2/2K}$ and
$\nolinebreak{K = v/u \equiv (1+V_0/\pi v)^{-1/2}}$ is the
conventional dimensionless 
parameter characterizing the LL interaction strength. Evaluation
of the $\omega$ integral leads to
\begin{equation}
\label{J8d}
I^{(\eta)}_{\theta\bar{\theta}}(t)=\gamma\arctan\left(1+t^2\Lambda^2\right)\,.
\end{equation}
At equilibrium 
$\int (dq)\langle \theta \theta\rangle_{\omega q}^{(\eta)}= 
(\pi\gamma/\omega)B_{\rm eq}(\omega)$. 
For $T=0$ this yields
\begin{eqnarray}
\label{J8dd}
I^{(\eta)}_{\theta\theta}(t)=\frac{\gamma}{2}\log\left(1+t^2\Lambda^2\right)\,.
\end{eqnarray}
Substituting Eqs.~(\ref{J8dd}), (\ref{J8d}) into Eq.(\ref{J6}), 
we reproduce the famous power-law behavior of TDOS, 
\begin{eqnarray}
\nu_{\rm eq}(\epsilon) \sim \nu_0  
\left(\epsilon/\Lambda\right)^\gamma \,.
\end{eqnarray}
At finite $T$ the long-time behavior of
$I^{(\eta)}_{\theta\theta}$ is modified,  
\begin{eqnarray}
I^{(\eta)}_{\theta\theta}(t) \simeq \pi\gamma Tt \equiv t/2\tau_\phi\,.
\end{eqnarray}
The ZBA dephasing rate is thus
$1/\tau_\phi= 2\pi\gamma T$ and 
agrees with the one relevant to 
Aharonov-Bohm effect \cite{lehur,GMP}.
While $1/\tau_\phi$ contributes to smearing of $\nu_{\rm eq}(\epsilon)$,
for $\gamma \lesssim 1$ this is not particularly important, 
as the singularity is anyway smeared on the
scale of  $T$ due to the distribution function $n_\eta(t)$ in
Eq.~(\ref{J6}). 

We turn now to the non-equilibrium situation. It is important to
emphasize that the functional bosonization procedure 
can be also applied to a (clean) LL with spatially dependent
interaction constant $K(x)$ (such a model was considered in
\cite{Maslov}). Since the interaction can still be gauged out,
the conclusion about independence of 
the fermionic distribution  $n_\eta(\epsilon)$
on $x$ retains its validity. 
On the other hand, the boson sector will be, most generally, 
characterized  by  different  
distribution functions $B_\eta^u(\omega)$, $B^v_\eta(\omega)$
corresponding to the modes propagating with velocities $u$ and $v$ 
[coexistence of such modes is clear from the correlation function
(\ref{J8})]. While the $u$-excitations are conventional plasmons, the
$v$-mode describes bare electron-hole pairs (thus moving with velocity $v$
of non-interacting fermions). The corresponding formalism for
higher-dimensional diffusive system was developed in \cite{Aleiner}
(where the analog of the $v$-mode was termed ``ghosts''); for an
analysis of energy relaxation in disordered LL in this framework, see
Ref.~\cite{Bagrets}.   We find
\begin{equation}
\int (dq)
\langle \theta \theta \rangle^R_{\omega,q}= 
\frac{\pi}{2\omega}
[\gamma B_L^u(\omega)
+ (\gamma+2) B_R^u(\omega)-2B_R^v(\omega)] 
\label{theta-theta} \nonumber
\end{equation}
and analogously for $\langle\theta\theta\rangle^L$. 

While the distribution $B_\eta^v$ of bare electron-hole pairs 
is determined by the fermion
distribution function $n_\eta$, see Eq.~(\ref{e7}), the plasmon
distribution $B_\eta^u$ may differ, depending on the setup.  
We consider first the ``adiabatic'' situation when
$K(x)$ changes slowly on the scale $l_\phi$ determined below. 
Then plasmons  with relevant wave vectors  are not backscattered by 
modulation of the interaction $K(x)$, so that 
$B^u_\eta=B^v_\eta\equiv B_\eta$ is given  by Eq.~(\ref{e7}). In this
situation Eq.~(\ref{theta-theta}) reduces to
\begin{equation}
\label{theta-theta-adiab}
\int (dq)
\langle \theta \theta \rangle^R_{\omega,q}= 
\pi\gamma [B_R(\omega)+B_L(\omega)]/2\omega\:.
\end{equation}
For the setup of Fig.~\ref{fig1}a Eq.~(\ref{theta-theta-adiab}) yields
$I^{(\eta)}_{\theta\theta} = \pi\gamma(T_L+T_R)t/2$. 
Therefore, the ZBA dips in TDOS of both chiral sectors 
(separated by $\mu_L-\mu_R = eV$) get
broadened (in addition to the thermal smearing) by the dephasing rate  
\begin{equation} 
1/\tau_\phi=\pi\gamma(T_L+T_R)\,.
\label{dephasing-setup-a}
\end{equation}
If one of $T_{R,L}$ is much lower than the other, so that
e.g. $T_L \ll 1/\tau_\phi$, the broadening of the corresponding ZBA minimum 
(near $\mu_L$) is determined by the dephasing.

\begin{figure}[t]
\includegraphics[width=0.4 \textwidth,angle=0]{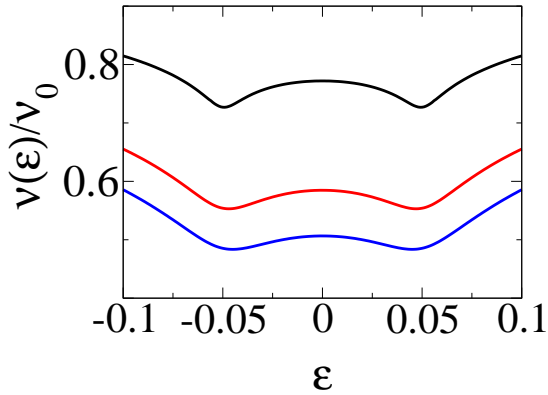}
\caption{ZBA in LL, setups b,c with $a=\frac{1}{2}$, $\Lambda=1$,
$T_\eta=0$,  $eV=0.1$, for $\gamma=$0.1,
  0.2, 0.25 (from top to bottom).} 
\label{fig2}
\end{figure}

For the ``fully non-equilibrium'' setups of
Fig.~\ref{fig1}b,c
the fermionic distributions $n_\eta(\epsilon)$ are of a
double-step form, 
\begin{equation}
n_\eta(\epsilon)=a_\eta
n_0(\epsilon_-)+ (1-a_\eta)n_0(\epsilon_+) \,,
\label{double-step}
\end{equation}
where $n_0(\epsilon)$ is the zero-$T$ Fermi
distribution (we assume $T_\eta\ll eV$), $0{<}a_\eta{<}1$,
and $\epsilon_\pm = \epsilon \pm V/2$ .
For the distribution (\ref{double-step}), one finds the  bosonic
distribution function 
\begin{eqnarray*}
&& \omega B^v_\eta(\omega)= [a_\eta^2+(1-a_\eta)^2 ] \omega B_{\rm eq}(\omega)
 +a_\eta(1-a_\eta) \\
&&
\times[(\omega+eV)B_{\rm eq}(\omega+eV) + (\omega-eV)B_{\rm eq}(\omega-eV)]\,.
\end{eqnarray*}
This yields the TDOS (see Fig.~\ref{fig2})
\begin{eqnarray}
\label{results}
\nu_\eta(\epsilon) &\simeq & 
a_\eta\nu_{{\rm eq}} ({\rm max}\{\epsilon_-, 
1/2\tau^{(\eta)}_\phi\}) \nonumber\\ 
&+&
(1-a_\eta)\nu_{\rm eq} ({\rm max}\{\epsilon_+, 
1/2\tau^{(\eta)}_\phi\})\,,
\end{eqnarray}
with the non-equilibrium  ZBA  dephasing rate
\begin{equation}
1/\tau^\eta_\phi=c_\eta\pi\gamma eV
\label{dephasing-double-step}
\end{equation}
and the numerical  
prefactor $c_\eta= a_R(1-a_R)+a_L(1-a_L)$.

If $K(x)$ varies fast, the plasmon distribution $B_\eta^u$
becomes spatially dependent,  while $B_\eta^v$ remains unchanged.
In the limit when the interaction is turned on as a sharp (on the scale
$l_\phi)$ step, the plasmons are scattered accordingly to the Fresnel
law \cite{Maslov}, with a reflection coefficient $R{=}(1-K)^2/(1+K)^2$. 
This yields a boundary condition for the distributions
$B_\eta^{u}$; e.g. for a sharp change on the r.h.s. of the wire
$ B_L^u(\omega){=} (1-R)B_L^v(\omega)+ R B_R^u(\omega)$,
and similarly for the left boundary. 
In this way it is easy to treat the situation with sharp switching of
the interaction on one or both sides \cite{unpublished}. 
The results (\ref{results}), (\ref{dephasing-double-step}) 
retain their validity but with modified numerical prefactors $c_\eta$.

To summarize, we have analyzed several setups in which non-equilibrium
LL can be observed. Using the functional bosonization formalism, we
have developed a theory of tunneling into such systems. While 
energy relaxation is absent, the dips of split ZBA are broadened by
dephasing, Eqs.~(\ref{dephasing-setup-a}), (\ref{dephasing-double-step}),
yielding a finite  quasiparticle life time. 
We reiterate that RPA  is exact for the setup
of Fig.~\ref{fig1}a,  but not for the setups of Figs.~\ref{fig1}b,c. 
In particular, the RPA value of the prefactor
$c_\eta$ in Eq.~(\ref{dephasing-double-step}) for $1/\tau_\phi$ should
be considered as an approximation. One can expect that RPA becomes
controllable for weak interaction, $\gamma\ll 1$.
Also, it remains to be seen whether 
exact results can be obtained for a generic non-equilibrium setup. 
The key observation is that the sum of vacuum diagrams ($\log Z_\phi$)
can be cast in the form  analogous to the generating function in 
the counting statistics problem \cite{Levitov-noise,Abanin}.
Work in these directions is underway.



We thank  
D. Bagrets, N. Birge,  A. Finkelstein, I. Gornyi, Y. Levinson, D. Maslov,
Y. Nazarov, D. Polyakov  
for useful discussions. This work was supported by
NSF-DMR-0308377 (DG), 
 US-Israel BSF, Minerva Foundation, and DFG SPP 1285 (YG),
EU Transnational Access Program RITA-CT-2003-506095
 (ADM), and Einstein Minerva Center.


\vspace*{-0.4cm}

\begin{thebibliography}{99}

\vspace*{-0.4cm}


\bibitem{gnt}M. Stone, \textit{Bosonization} (World Scientific, 1994);
A.O. Gogolin, A.A. Nersesyan, and A.M. Tsvelik,
\textit{Bo\-son\-iza\-tion in Strongly Correlated Systems},
(University Press, Cambridge 1998);
T. Giamarchi, \textit{Quantum Physics in One Dimension},  (Claverdon Press Oxford, 2004).



\bibitem{zba-carbon-nanotubes}
 M. Bockrath,{\it et al}., Nature (London) {\bf 397}, 598 (1999); Z. Yao {\it et al.}, Nature (London){\bf 402}, 273 (1999).


\bibitem{Saleur}
P. Fendley, A.W.W. Ludwig, H. Saleur,
Phys. Rev. Lett. {\bf 75} 2196, (1995);
C.D. Chamon, D.E. Freed, X.G. Wen, Phys. Rev. B  {\bf 53}, 4033 (1996).


\bibitem{Feldman} 
D.E. Feldman, Y.  Gefen,  Phys. Rev. B {\bf 67}  115337 (2003).

\bibitem{energy_relaxation} D.C. Mattis and E.H. Lieb, J. Math. Phys. 
{\bf 6}, 375 (1965); M. Khodas {\it et al.}, 
Phys. Rev. B {\bf 76}, 155402 (2007). 


\bibitem{Pothier} A. Anthore {\it et al.},  
Phys. Rev. Lett. {\bf 90}, 076806 (2003).

\bibitem{Chudnovsky} The difference in temperatures  distinguishes
this setup from that of M. Trushin, A. L. Chudnovskiy,
arXiv:0705.4552,  where  $T_L=T_R$
(so that bosons are in equilibrium, and the usual Matsubara
bosonization technique can be applied). 


\bibitem{GGM} D.B. Gutman, Y. Gefen, A.D. Mirlin, Phys. Rev. Lett. {\bf 100}, 086801 (2008).


\bibitem{GMP} I. V. Gornyi, A. D. Mirlin, D. G. Polyakov, 
Phys. Rev. Lett. {\bf 95}, 046404 (2005); Phys. Rev. B{\bf 75}, 085421
(2007). 


\bibitem{lehur} K. Le Hur, Phys. Rev. Lett. {\bf 95}, 076801 (2005);
  Phys. Rev. B {\bf 74}, 165104 (2006). 
  
  
\bibitem{Fogedby:76}
H.~C. Fogedby, J. Phys. {\rm C} {\bf 9}, 3757 (1976).

\bibitem{LeeChen:88}
D.~K. Lee and Y.~Chen, J. Phys. {\rm A} {\bf 21}, 4155 (1988).




\bibitem{Yur} 
A.~Grishin, I.V. Yurkevich, and I.V.  Lerner,
Phys. Rev. B {\bf 69}, 165108 (2004); I.V. Lerner and I.V. Yurkevich,
in {\it 
Nanophysics: Coherence and Transport} (Elsevier, 2005), p.109.
 
\bibitem{DzLar:73}
I.~E. Dzyaloshinskii and A.~I. Larkin, Sov. Phys. JETP {\bf 38},
202 (1973).



\bibitem{Maslov} D.L. Maslov and M. Stone Phys. Rev. B {\bf 52}, R5539 (1995);
R. Fazio, F.W.J. Hekking, D.E. Khmelnitskii 
Phys. Rev.Lett. {\bf 80}, 5611 (1998).



\bibitem{Aleiner} G. Catelani and I.L. Aleiner, JETP {\bf 100}, 331,  (2005).

\bibitem{Bagrets} D.A. Bagrets, I.V. Gornyi, D.G. Polyakov, unpublished.

\bibitem{Levitov-noise} L.S. Levitov, H. Lee ,G.B. Lesovik 
J. Math. Phys. {\bf 37}, 4845 (1996);
L.S. Levitov, G.B. Lesovik, JETP Lett. {\bf 58}, 230 (1993).

\bibitem{Abanin} D. A. Abanin, L. S. Levitov,
Phys. Rev. Lett. {\bf 94}, 186803 (2005); I. Snyman, Y. V. Nazarov,
arXiv:0801.2293.

\bibitem{unpublished} D.B. Gutman, Y. Gefen and A.D. Mirlin,
to be published.





\end{thebibliography}
\end{document}